\documentclass[smallextended]{svjour3}                     
\smartqed  
\usepackage{graphicx}
\usepackage{amsmath}
\usepackage{amssymb}
\newcommand{\beq}{\begin{equation}}
\newcommand{\eeq}{\end{equation}}

\newcommand{\pder}[2]{\frac{\partial #1}{\partial #2}}

\newcommand{\gw}{\mathrm{gw}}

\newcommand{\F}{{\cal F}}
\newcommand{\C}{{\cal C}}

\DeclareMathOperator{\MeV}{MeV}
\DeclareMathOperator{\GeV}{GeV}

\DeclareMathOperator{\Hz}{Hz}

\begin{document}

\title{Signatures of the neutrino thermal history in the spectrum of primordial gravitational waves
}


\author{Riccardo Benini  \and  Massimiliano Lattanzi \and Giovanni Montani 
}


\institute{R. Benini \at
              ICRA and Physics Department, University of Rome, ``La Sapienza'', P.le Aldo Moro 2, 00185 Rome, Italy \\
           \and
           M. Lattanzi \at
              ICRA and Physics Department, University of Rome, ``La Sapienza'', P.le Aldo Moro 2, 00185 Rome, Italy \\
              \email{lattanzi@icra.it}
              \and
            G. Montani \at
            ENEA -- C.R. Frascati (Department F.P.N.),
Via Enrico Fermi, 45 (00044), Frascati (Rome), Italy\\
ICRA and Physics Department, University of Rome, ``La Sapienza'', P.le Aldo Moro 2, 00185 Rome, Italy \\
}

\date{Received: date / Accepted: date}

\maketitle

\begin{abstract}
In this paper we study the effect of the anisotropic stress generated by neutrinos on the propagation of primordial cosmological gravitational waves.
The presence of anisotropic stress, like the one generated by free-streaming neutrinos, partially absorbs the gravitational waves (GWs) propagating across the Universe. We find that in the standard case of three neutrino families,
22\% of the intensity of the wave is absorbed, in fair agreement with previous studies. We have also calculated the maximum possible amount of damping, corresponding to the case of a flat Universe completely dominated by ultrarelativistic collisionless particles. In this case 43\% of the intensity of the wave is absorbed. Finally, we have taken into account the effect of collisions, using a simple form for the collision term parameterized by the mean time between interactions, that allows to go smoothly from the case of a tigthly-coupled fluid to that of a collisionless gas. 
The dependence of the absorption on the neutrino energy density and on the effectiveness of the interactions opens the interesting possibility of observing spectral features related to particular events in the thermal history of the Universe, like neutrino decoupling and electron-positron annihilation, both occurring at $T\sim 1 \MeV$. GWs entering the horizon at that time will have today a frequency $\nu\sim 10^{-9} \Hz$, a region that is going to be probed by Pulsar Timing Arrays.

\end{abstract}

\section{Introduction}
\label{intro}

The presence in the Universe today of a stochastic background of gravitational waves (GWs) is a quite general prediction of several early cosmology scenarios. In fact, the production of gravitational waves is the outcome of many processes that could have occurred in the early phases of the cosmological evolution. Notable examples of this kind of processes include the amplification of vacuum fluctuations in inflationary \cite{Gr75} and pre-big-bang cosmology scenarios \cite{Ga93}, phase transitions \cite{Ho86}, and finally the oscillation of cosmic strings loops \cite{Vi81}. In most of these cases, the predicted spectrum of gravitational waves extends over a very large range of frequencies; for example, inflationary expansion produces a flat spectrum that spans more than 20 orders of magnitude in frequency, going from $10^{-18}$ to $10^{9}$ Hz.

The detection of such primordial gravitational waves, produced in the early Universe, would be a major breakthrough in cosmology and high energy physics. This is because gravitational waves decouple from the cosmological plasma at very early times, when the temperature of the Universe is of the order of the Planck energy. In this way, relic gravitational waves provide us a ``snapshot'' of the Universe near the Planck time, in a similar way as the cosmic microwave background  (CMB) radiation images the Universe at the time of recombination.

The extremely low frequency region ($\nu_0\lesssim10^{-15}\Hz$) in the spectrum of primordial gravitational waves can be probed through the anisotropies of the CMB. In particular, gravitational waves leave a distinct imprint in the so-called magnetic or B-modes of its polarization field \cite{Hu97,Pr05}. The amplitude of the primordial spectrum of gravitational waves is usually parameterized through the tensor-to-scalar ratio $r$, i.e., the ratio between the amplitudes of the initial spectra of the tensor and scalar perturbations in the metric. The Planck satellite \cite{Planck} is expected to be sensitive \cite{Tu05} to $r\ge0.05$, 
correspondingto a density parameter $\Omega_{\gw}(\nu)\equiv (1/\rho_c) d\rho_{\gw}/d\log\nu $ as faint as $\sim 3\times10^{-16}h^{-2}$ ($h$ is the dimensionless Hubble constant);
future experiments are expected to enhance the sensitivity of two orders of magnitude~\cite{Tu05}.

On the other hand the operating large-scale interferometric GW detectors, although designed with the aim to detect astrophysical signals, can possibly also detect signals of cosmological origin \cite{Ma00}. They give complementary information with respect to the CMB polarization field since they probe a different region in the frequency domain. In particular the ground-based interferometers, such as the LIGO \cite{LIGO}, VIRGO \cite{VIRGO}, GEO600 \cite{GEO600} and TAMA300 \cite{TAMA300} experiments, operate in the range between $~10\,\Hz$ and few kHz, and are expected to be sensitive to $\Omega_{\gw}h^2\ge 10^{-2}$. Even more interesting is the LISA space interferometer \cite{LISA}, that will hopefully operate in the 2020s. Not being hampered by the Earth seismic noise, it will probe the frequency region between $10^{-4}$ and $1\,\Hz$ and will in principle be able to detect $\Omega_\gw h^2\ge 10^{-12}$ at $\nu_0=10^{-3}\,\Hz$. According to theoretical predictions, a large enough GW signal at this frequencies can be produced, with the appropriate choice of parameters, by a pre-big-bang accelerated expansion, by the oscillation of cosmic strings, or by the electroweak phase transition occurring at $T=300\,\GeV$. Finally, pulsar observations can be used to obtain information on the stochastic GW background, through the technique known as pulsar timing. The so-called ``Pulsar Timing Arrays'' will probe the region around $10^{-9} \Hz$ \cite{PTA}.

In order to compare the theoretical predictions with the expected instrument sensitivities, one needs to evolve the GWs from the time of their production to the present. It is often assumed that gravitational waves propagate in vacuum, i.e., they freely stream across the Universe. In this case, the only effect on a propagating GW is a change in frequency (corresponding to the usual redshift of the 	wavelengths caused by the expansion of the Universe), and a corresponding change in the energy of the wave. However, GWs are sourced by the anisotropic stress part of the energy-momentum tensor of matter, so that the vacuum approximation is well-motivated only when this can be neglected. It is already known that the anisotropic stress of free streaming neutrinos acts as an effective viscosity, absorbing gravitational waves in the low frequency region, thus resulting in a damping of the B-modes of CMB \cite{Bo96,Du98,We04,Xi08,Zh09}. 

In the present work we aim to start a study concerning the possible effects of the presence of neutrinos in other frequency ranges,
like those probed by Pulsar Timing Arrays or interferometers. In particular, we aim to understand if events occurring in the thermal history of the Universe, like neutrino decoupling or the electron-positron annihilation,
can leave an imprint in the spectrum of cosmological gravitational waves. The rationale behind this is that these events give rise to sharp changes in the neutrino density and in the neutrino mean free path; corresponding in turn to sharp changes in the anisotropic stress of the cosmological fluid. This would point to the fact that GWs entering the horizon before or after these events would experience a different amount of absorption.

The paper is organized as follows. In Sec. 2, we introduce the basic equations. In particular we introduce the coupled Einstein-Boltzmann system and recast it in a form that is very suitable for numerical integration.
In Sec. 3, we show the results of the numerical integration of the Einstein-Boltzmann system, showing the time dependence of the GW amplitude and computing the amount of absorption for different values of the neutrino density and of the neutrino mean free-path. Finally, in Sec. 4 we draw our conclusions and put forward some ideas for the future.

\section{Basic equations}
\label{sec:prop}

We shall use, all throughout the paper, natural units in which $c=\hbar=k_B=1$.

Let us consider a gravitational wave, propagating on the background of a flat Friedmann Universe. In synchronous gauge, the spatial components of the perturbed metric are written as
\footnote{We use the $(-+++)$ signature for the metric.}:
\begin{equation}
g_{ij}=a^2(t)\left[\delta_{ij}+h_{ij}\right]
\end{equation}
while the other components are left unperturbed: $g_{00}=-1$ and $g_{0i}=0$. We will consider only the transverse traceless part of $h_{ij}$, representing 
a GW. Here $a(t)$ is the cosmological scale factor, that evolves according to the background Friedmann equation:
\begin{equation}
\left(\frac{da}{dt}\right)^2=\frac{8\pi G}{3} a^2 \bar\rho
\label{eq:fried}
\end{equation}
where $\bar \rho$ is the background density of the cosmological fluid (in general, we will use overbars to denote background quantities).

The components of the tensor $h_{ij}$ evolve according to \cite{We72}:
\begin{equation}
\partial^2_t{h}_{ij}+\left(\frac{3}{a}\frac{da}{dt}\right)\partial_t{h}_{ij}-\left
(\frac{\nabla^2}{a^2}\right)h_{ij}=16\pi G\frac{\Pi_{ij}}{a^2}\;,
\label{eq:einst}
\end{equation}
where $\Pi_{ij}$ is the anisotropic stress, i.e. the traceless part of the three dimensional energy-momentum tensor $T^{i}_{j}$ of the cosmological fluid. It is then defined through the relation $\Pi^i_j=T^i_{ j}-\delta^i_j T_k^k/3=T^i_j-{\cal P}\delta^i_{j}$, where $\cal{P}$ is the total pressure of the fluid (including possibly a small perturbation with respect to the background).

The macroscopic properties of the cosmological fluid can be derived by the phase space distribution of its particles. The phase space is described by three positions $x^i$ and by their three conjugate momenta $P_i\equiv mdx_i/ds$. The proper momentum $p_i=p^i$ measured by an observer at a fixed spatial coordinate is related to $P_i$ by $P_i=a(\delta_{ij}+\frac{1}{2}h_{ij})p^j$. As usual, the phase-space distribution function (DF) $f(x^i, P_j,t)$ of the particles gives the number of particles inside the 6-dimensional volume element:
\begin{equation}
f(x^i, P_j,t) d^3x d^3P = dN.
\end{equation}

The energy-momentum tensor can be written in terms of the distribution function as follows:
\begin{equation}
T^\mu_\nu=\frac{1}{\sqrt{-g}}\int f(x^i, P_j,t) \frac{ P^\mu P_\nu}{P^0}\,dP_1\,dP_2\,dP_3,
\end{equation}
where $g$ is the determinant of the metric.

In the following, we will use the comoving three-momentum $q_i = a p_i$ in place of $P_i$ as a momentum variable, and write it as $q_i=q n_i$ where $n_i$ is a unit vector. We will also use the conformal time $\eta$, defined by  $dt= a\, d\eta$ as our time variable. Then we write $f=f(x^i,q,n_j,\eta)$. Finally, we write the DF as the sum of a zeroth-order, unperturbed part $f_0$, and a small perturbation:
\begin{equation}
f(x^i,q,n_j,\eta)=f_0(q)\left[1+\Psi(x^i,q,n_j,\tau)\right].
\end{equation}
Using the fact that $\sqrt{-g}=a^4/(1-\frac{1}{2}h)$ and $dP_1 dP_2 dP_3=(1+\frac{1}{2} h) q^2 dq d\Omega$, where $h$ is the trace of $h_{ij}$ and $d\Omega$ is the infinitesimal element of solid angle around $\hat n$,
we can write:
\begin{equation}
T^i_j = a^{-4} \int \frac{q^2}{\epsilon}  n^i n_jf_0(q)(1+\Psi) q^2 dq d\Omega
\label{eq:Tij}
\end{equation}
where $\epsilon\equiv\sqrt{q^2+a^2m^2}$. We warn the reader that, even if we follow the convention of distinguishing between covariant and contravariant indices, the fact that we
use the flat metric $\delta_{ij}$ to raise and lower the indices of $p_i$ (and hence $n_i$) means that equalities like that in Eq. (\ref{eq:Tij}) are not covariant.
The unperturbed phase space distribution is given by a thermal equilibrium distribution, i.e. by a Fermi-Dirac or Bose-Einstein distribution:
\begin{equation}
f_0(q)=\frac{g_s}{(2\pi)^3}\frac{1}{e^{\epsilon/T_0}\pm 1},
\end{equation}
where $g_s$ is the number of quantum degrees of freedom,  and $T_0$ is the present temperature of the particles.

The DF evolves according to the Boltzmann equation:
\begin{equation}
\hat L[f]=\hat C[f]
\label{eq:boltzeq}
\end{equation}
where the $\hat L\equiv Df/D\eta$ is the Liouville operator, and $\hat C$ is the collision operator accounting for collisions between particles. Using the geodesic equation, the  Liouville operator $\hat L$ can be cast in the form (to first order in perturbed quantities):
\begin{equation}
\hat L[f]\equiv\frac{Df}{D\eta}=\pder{f}{\eta}+\frac{dx^i}{d\eta}\pder{f}{x^i} 
-\frac{1}{2}q n^i n^j\frac{d h_{ij}}{d\eta}\pder{f}{q}.
\label{eq:liouv}
\end{equation} 
Once the collision term is also specified, Eqs. (\ref{eq:fried}), (\ref{eq:einst}), (\ref{eq:Tij}), (\ref{eq:boltzeq}) and (\ref{eq:liouv}) are all that is needed, at least in principle, to follow the propagation of a GW.

\subsection{Multipole formalism}

In this subsection, we will rewrite the coupled Einstein-Boltzmann system derived above to a form that is more suitable for numerical integration.
With very small variations, this is the same procedure used when dealing with scalar perturbations \cite{Ma:1995ey}.
First of all, we note that we will only be concerned with massless particles as a source for the anisotropic stress,
so we will set the mass $m$ equal to zero in all the formulas derived above. Although we will be referring to these particles as ``neutrinos'',
for the purpose of computing the evolution of cosmological GWs they could actually be everything as long as they are effectively massless, i.e. as long as the temperature of the cosmological plasma
is much larger than their rest mass.

Firs of all, we Fourier transform the spatial dependence of all the relevant quantities introduced above. With a slight abuse of notation,
we shall use the same symbol to denote a given quantity and its Fourier transform. The Boltzmann equation in $k$-space reads (dots denote 
derivatives with respect to conformal time):
\begin{equation}
\dot\Psi +i k_i n^i\Psi-\frac{1}{2}n^i n^j\dot h_{ij}\frac{d\ln f_0}{d \ln q} = \frac{1}{f_0} \hat C[f]
\end{equation}
In the case of massless particles, the dependence of the DF from $q$ can be integrated out. In particular,
after defining
\begin{equation}
F_\nu(k_i,\, n_j,\,\tau) \equiv \frac{\int q^3 f_0(q) \Psi(k_i,\,q,\, n_j,\,\tau) dq}{\int q^3 f_0(q) dq},
\end{equation}
we can multiply the Boltzmann equation by $q^3 f_0$ and integrate over $q$; the result is:
\begin{equation}
\dot F_\nu + i\, k_i n^i F_\nu + 2\dot h_{ij}n^i n^j = \frac{4\pi}{a^4 \bar \rho_\nu}\int q^3 \hat C[f] dq, 
\label{eq:boltzF}
\end{equation}
where we have also used the fact that $\bar\rho_\nu=4\pi a^{-4}\int q^3 f_0(q) dq$. Then we define:
\begin{equation}
\F_{ij}(k_i,\,\mu,\,\tau) = \int_0^{2\pi} \left(n_i n_j-\frac{\delta_{ij}}{3}\right) F_\nu\, d\phi,
\end{equation}
where $\phi$ is the polar angle, so that the infinitesimal solid angle element $d\Omega = \sin\theta d\theta\,d\phi$. 
Multiplying Eq. (\ref{eq:boltzF}) by $(n_i n_j-\delta_{ij}/3)$ and integrating over $\phi$, we get (we define as usual $\mu \equiv \hat k \cdot \hat n$):
\begin{equation}
\dot \F_{ij} + i k \mu\, \F_{ij} +2\dot h_{l m}\int_0^{2\pi} n^l n^m \left(n_i n_j-\frac{\delta_{ij}}{3}\right) d\phi = {\cal C}_{ij}
\label{eq:boltz2}
\end{equation}
where we have defined the ``collision term'' ${\cal C}_{ij}$:
\begin{equation}
{\cal C}_{ij}\equiv\frac{4\pi}{a^4 \bar \rho}\int q^3 dq \int_0^{2\pi} d\phi \, \left(n_i n_j -\frac{\delta_{ij}}{3}\right) \hat C[f], 
\end{equation}
Now we expand $\F_{ij}$ and ${\cal C}_{ij}$ in Legendre polynomials:
\begin{align}
\F_{ij}(k_i,\,\mu,\,\tau) &= \sum_{\ell=0}^\infty (-i)^{\ell}(2\ell+1)\F_{ij}^{(\ell)}(k_i,\,\tau)P_{\ell}(\mu)\\
\C_{ij}(k_i,\,\mu,\,\tau) &= \sum_{\ell=0}^\infty (-i)^{\ell}(2\ell+1)\C_{ij}^{(\ell)}(k_i,\,\tau)P_{\ell}(\mu)
\end{align}
In order to obtain the ``tower'' of (infinite) differential equations for the $\F_{in}^{(\ell)}$, we multiply Eq. (\ref{eq:boltz2}) by $(i^\ell/2) P_{\ell}$, integrate over $\mu$ and use the 
orthogonality relation of the Legendre polynomials, i.e: 
\begin{equation}
\int_{-1}^{1} P_{\ell} P_{m}\,d\mu = \frac{2}{2\ell+1} \delta_{\ell m}
\end{equation}
The detailed calculation is shown in the appendix. The final result is:
\begin{align}
&\dot \F_{ij}^{(0)} = - k\,\F_{ij}^{(1)} - \frac{8\pi }{15} \dot h_{ij } +\C_{ij}^{(0)}, \label{eq:bolt_mpol1} \\[0.5cm]
&\dot \F_{ij}^{(2)} = -\frac{k}{5} \left[   3 \F_{ij}^{(3)} -  2 \F_{ij}^{(1)} \right] - \frac{16\pi}{105} \dot h_{ij }+\C_{ij}^{(2)},\\[0.5cm]
&\dot \F_{ij}^{(4)} = - \frac{k}{9} \left[   5 \F_{ij}^{(5)} -  4 \F_{ij}^{(3)} \right] - \frac{8\pi}{315} \dot h_{ij } +\C_{ij}^{(4)},\\[0.5cm]
&\dot \F_{ij}^{(\ell)} = -\frac{k}{2\ell+1} \left[   (\ell+1) \F_{ij}^{(\ell+1)} -  \ell\, \F_{ij}^{(\ell-1)} \right] +\C_{ij}^{(\ell)}\qquad (\ell \ne 0,2,4). \label{eq:bolt_mpol4}
\end{align} 
This system of infinite first-order ordinary differential equations is completely equivalent to the original Boltzmann equation and can be solved
with fairly standard numerical methods for ODEs.

The system should be closed with the evolution equation for $h_{ij}$, namely Eq. (\ref{eq:einst}). In Fourier space,
and using conformal time, this reads:
\begin{equation}
\ddot h_{ij} + 2{\cal H} \dot h_{ij} + k^2 h_{ij} =16\pi G \Pi_{ij},
\label{eq:einstK}
\end{equation}
where ${\cal H}$ is the ``conformal'' Hubble constant ${\cal H} = \dot a/a$. The anisotropic stress $\Pi_{ij}$ is given by:
\begin{equation}
\Pi_{ij}=T_{ij}-\frac{g_{ij}}{3}T^k_k =
\frac{a^2\bar\rho}{4\pi}\int  \left(n_i n_j -\frac{\delta_{ij}}{3}\right) F_\nu d\Omega 
= \frac{a^2\bar\rho_\nu}{4\pi}\F_{ij}^{(0)}.
\end{equation}
In deriving this expression we have used the fact that $\int n_i n_j d\Omega=4\pi \delta_{ij}/3$. Then we finally get:
\begin{equation}
\ddot h_{ij} + 2{\cal H} \dot h_{ij} + k^2 h_{ij} =4 G a^2 \bar \rho_\nu \mathcal{F}_{ij}^{(0)}.
\end{equation}

Finally, we have to specify the initial conditions for the integration. By studying the behaviour of the solution when the wave is far outside the horizon ($k\eta \gg 1$) it can be seen that the 
right initial conditions are $\dot h_{ij} =0$ and $\F_{ij}^{(\ell)}=0$. The initial value $h_{ij}^{(0)}$ of $h_{ij}$ is arbitrary but the equations can always be rescaled to have $h_{ij}^{(0)}=1$.

\section{Interaction of gravitational waves with neutrinos}

We can now use the equations derived in the previous section to study the effect of neutrinos 
on the propagation of cosmological GWs. We restrict our attention to waves entering the horizon well
before the time of matter-radiation equality, corresponding to a redshift $z\simeq 10^4$. This corresponds
to waves with a present frequency $\nu \gg 10^{-16}$ Hz. During the radiation-dominated era, $a\propto \eta$ and ${\cal H}=1/\eta$.
It can be shown that using the time variable $u\equiv k\eta$ the evolution equations can be recast in a form such that 
$k$ does not explicitly appear, so that the evolution in with respect to $u$ is independent from the frequency of the wave. Also,
the convenience of using $u$ is that $u=k\eta \sim 1$ corresponds to the time of horizon crossing. The results of the numerical integration
should be compared with the solution in the absence of anisotropic stress ($\Pi_{ij} = 0$), i.e. $h_{ij} = h_{ij}^{(0)} \sin u/u$.

\begin{center}
\begin{figure}[th]
\includegraphics[width=0.8\textwidth]{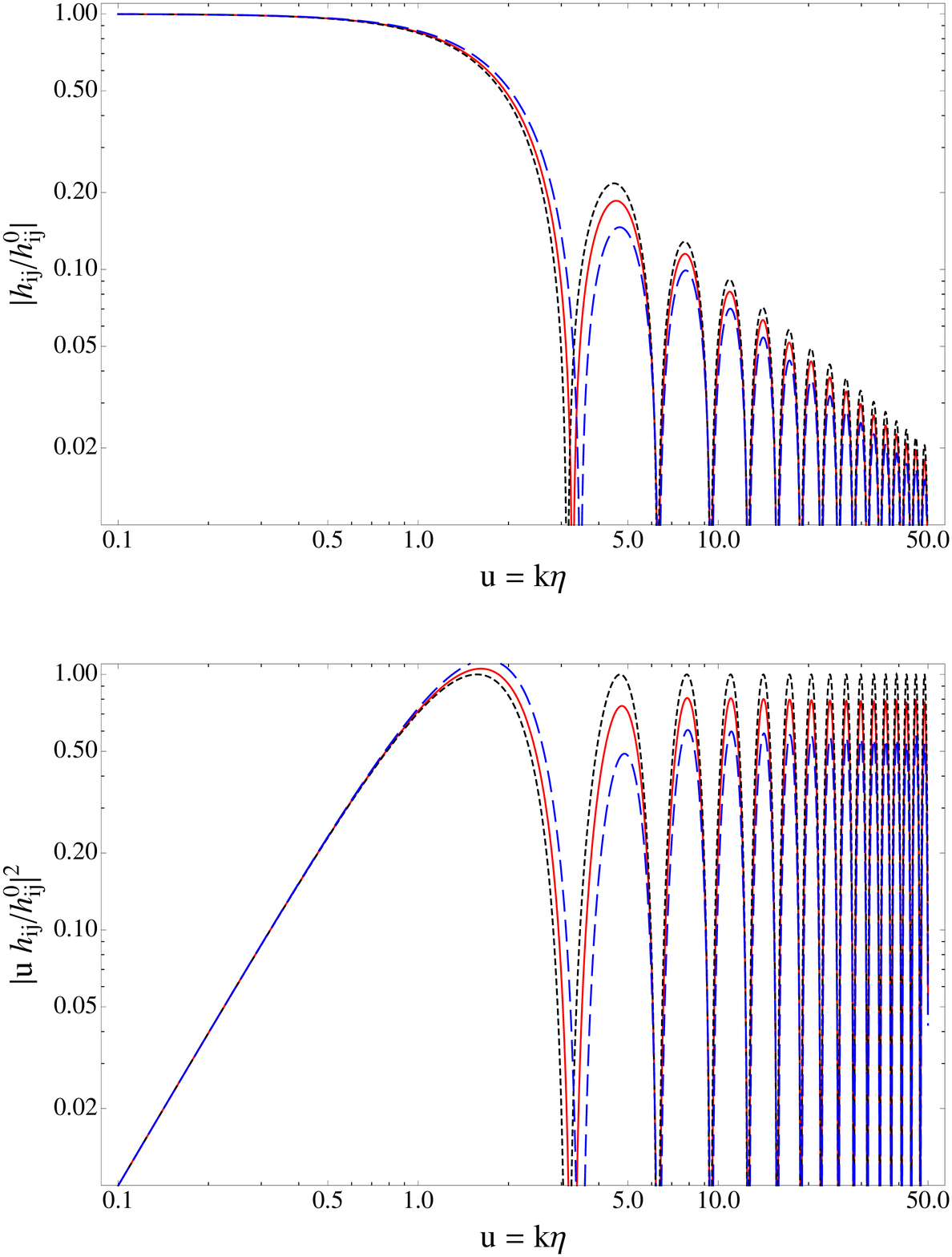}
\caption{Top panel: Evolution of the normalized wave amplitude $h_{ij}/h_{ij}^{(0)}$ for $f_\nu=0.4$ (red solid curve), $1$ (blue dashed curve), $0$ (black dotted curve). Bottom panel:
The same as the top panel, but for the normalized wave intensity $|h_{ij}/h_{ij}^{(0)}|^2$ times $u^2$, so that the redshift due to the expansion has been taken out.}
\label{fig:hvsnu_fnu04}
\end{figure}
\end{center}

First we consider the case of a vanishing collision term, $\C_{ij}^{(\ell)}=0$. This is the case after neutrino
decoupling, occurring when the temperature of the Universe is $T\simeq 1~\MeV$ and $1+z\sim 10^{10}$. At lower
temperatures, neutrinos do not interact with the other particles in the cosmological plasma so that they are
free-streaming and collisions are effectively absent. This is basically the case that was considered in Ref. \cite{We04}.
In order to parameterize the neutrino density, it is useful to introduce the quantities $R_\nu \equiv \bar\rho_\nu/\bar\rho_\gamma$ and  $f_\nu\equiv \bar \rho_\nu /\bar \rho$,
related by $f_\nu = R_\nu/(1+R_\nu)$. 
Taking the standard case of three neutrino families with a temperature $T_\nu = (4/11)^{1/3} T_\gamma$, we have that:
\begin{equation}
\bar \rho_\nu=3\times\frac{7}{8}\left(\frac{4}{11}\right)^{1/3}\bar \rho_\gamma.
\end{equation}
so that $R_\nu = 0.6813$ and $f_\nu =0.4052$. The results of the numerical integration performed using this value of $f_\nu$ is shown in Fig. \ref{fig:hvsnu_fnu04}.
In the top panel we plot the evolution of $h_{ij}$ (divided by its initial value $h_{ij}^{(0)}$) with respect to $u$ (red curve). We see that, as it should be expected, the amplitude $h_{ij}$ is constant
outside the horizon and starts decreasing after the horizon crossing; this is mainly due to the redshift caused by the expansion of the Universe. However, when comparing with
the zero-stress solution $\sin u/u$ (black short-dashed curve) it can be noticed that the wave suffers an additional damping, caused by the anisotropic stress of the neutrinos. This is made
even more clear in the bottom panel, where we plot the combination $u^2 |h_{ij}|^2$ in order to see the behaviour of the intensity $|h_{ij}|^2$ once the expansion of the Universe has been taken out.
We se that the amount of damping with respect to the zero-stress case tends to a constant value. For $f_\nu=0.4$, the intensity of the wave is $0.78$ times its value in the absence of stress, 
so that roughly 22\% of the intensity of the wave is absorbed. 
The amount of absorption increases with the neutrino fraction $f_\nu$. It is worth stressing, at this point, that the cosmological
neutrino background has not been directly observed yet. Although strong deviations from the standard scenario with $f_\nu=0.4$ are unlikely, the possibility that $f_\nu$ has a different value
(due for example to the presence of additional particles in the early Universe) should be taken into account. For this reason we show in Fig. \ref{fig:hvsnu_fnu04} also the extreme case $f_\nu=1$ (blue long dashed curve), i.e. $\bar \rho_\nu = \bar \rho$, meaning that all the matter content of the Universe is made of non-interacting, ultrarelativistic particles. This gives the maximum possible amount of damping: the intensity of the wave is nearly halved, being $0.57$ times its zero-stress value.

\begin{center}
\begin{figure}[th]
\includegraphics[width=0.8\textwidth]{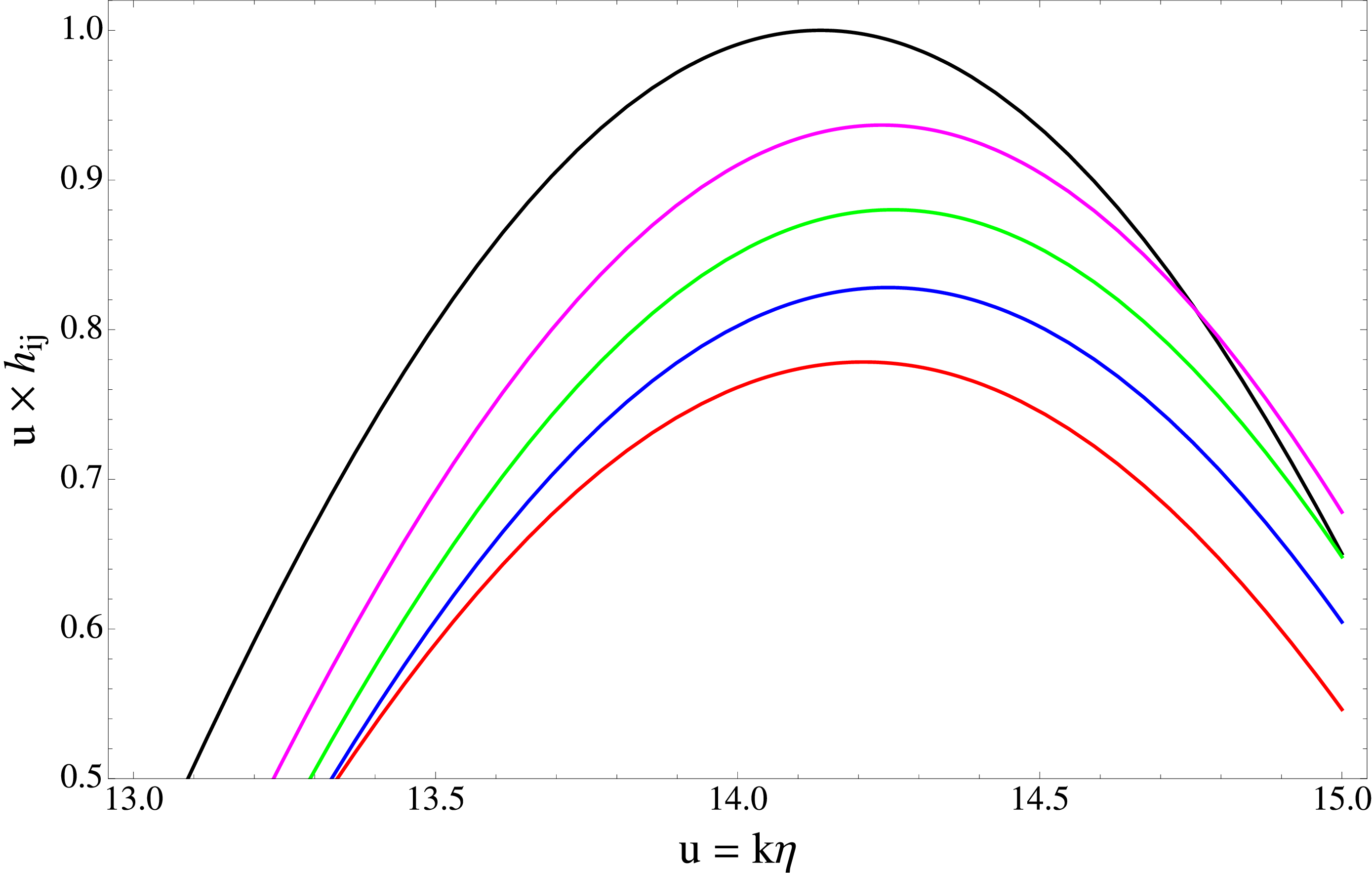}
\caption{Evolution of the wave for $f_nu = 1$ and different value of the mean collision time $\tau$. From top to bottom: $k\tau = 0.5,\,1,\,2,\,10$.}
\label{fig:hvsu_int}
\end{figure}
\end{center}

Then we turn to consider the effect of collisions. The exact computation of the collision term depends on the details of the interaction. However, a useful although rough approximation consists
in writing $\hat C[f] = - f_0\Psi /\tau$, where $\tau$ is the average time between collisions. This will give $\C_{ij}^{(\ell)}=-\F_{ij}^{(\ell)}/\tau$. The key parameter in defining the strength of the interactions is $k \tau$, that is basically the ratio of the frequency of the wave to the frequency of the collisions. A very small value of $k \tau$ would correspond to very frequent and thus effective collisions; the right-hand sides of Eqs. 
(\ref{eq:bolt_mpol1})-(\ref{eq:bolt_mpol4}) would be dominated by the collision terms and would have the solution $\F_{ij}\propto e^{-\eta/\tau}$, meaning that the anisotropic stress would decay exponentially. On the other hand, a large value of $k \tau$ would correspond to rare collisions and in the limit $k \tau\to\infty$ the collisionless result should be recovered. Without resorting to any particular model, we show in Fig. \ref{fig:hvsu_int} the result of the numerical integration for different constant values of $k \tau$ ranging from 0.5 to 10. We fix the neutrino fraction to the extreme value $f_\nu =1 $ in order to make the differences more evident. We see from the figure that the smaller 
the value of $k\tau$, the more the amplitude of the wave tends to its undamped value of 1, while when the collisions are rare (large $k\tau$) we recover the large damping found above.

\section{Conclusion and prospects}

In this previous sections we have studied the effect of the anisotropic stress generated by massless particles (``neutrinos'') on the propagation of cosmological gravitational waves.
In particular, we have put the relevant equations in a form that is very suitable for numerical integration and allows a quite clear quantitive understanding of the effect of the relevant parameters.
The presence of anisotropic stress, like the one generated by free-streaming neutrinos, partially absorbs the GWs propagating across the Universe. In the standard case of three neutrino families,
the wave is damped by a factor 0.78 in intensity (0.88 in amplitude). We have also calculated the maximum possible amount of damping, corresponding to the case of a flat Universe completely
dominated by ultrarelativistic collisionless particles. In this case the wave is damped by a factor 0.57 in intensity (0.75 in amplitude). Finally, we have taken into account the effect of collisions, and,
using a simple form for the collision term  parameterized by the mean time between collisions, we have shown that we can go smoothly from the case of a tigthly-coupled fluid to that of a collisionless fluid.

The dependence of the amount of damping from the neutrino fraction and from the effectiveness of the interactions opens interesting possibility. In fact, neither one of these quantities is really constant 
during the history of the Universe. The density of ultrarelativistic particles experiences several abrupt changes during the cosmological evolution, corresponding to the
creation of new particle species when the temperature of the Universe is large enough. The effectiveness of the interactions depends on the number density of target particles and on the interaction cross section, and both these quantities are a function of the temperature. Since the evolution of a gravitational wave is mainly affected by events occurring around the time of its horizon entry, and this in turn is directly related to the wave frequency, this open the interesting possibility of observing spectral features related to particular events in the thermal history of the Universe. A very promising frequency range is the nanohertz range. This roughly corresponds to GWs entering the horizon when $T\sim 1 \MeV$, corresponding not just to one, but to two notable events in the thermal history of the Universe: neutrino decoupling and electron-positron annihilation.
The first corresponds to the transition, for the neutrinos, from being a tigthly coupled fluid to being a collisionless gas, so it can be roughly thought as representing a (quite fast) change in the value of $\tau$. The second event, $e^+e^-$ annihilation, occuring shortly thereafter, changes the ratio of the photon and neutrino temperatures, and thus marks a sudden change in the value of $f_\nu$ from $0.72$ to $0.40$. Interestingly, the nanoHertz frequency region is going to be probed by the so-called Pulsar Timing Arrays \cite{PTA} and thus represents a very promising observational target possibly allowing to increase our knowledge of the thermal history of the early Universe.

\section*{Appendix}
In this appendix we show the calculations leading to Eqs. (\ref{eq:bolt_mpol1})-(\ref{eq:bolt_mpol4}) from  Eq. (\ref{eq:boltz2}).
Let us consider separately the three terms in Eq. (\ref{eq:boltz2}).
\paragraph{First term.}
This is simply:
\begin{equation}
\frac{i^\ell}{2}\int_{-1}^{1} d\mu \dot \F_{ij} P_{\ell} = \dot \F_{ij}^{(\ell)}
\end{equation}
\paragraph{Second term.}
We need to calculate
\begin{equation}
\frac{i^\ell}{2} ik \int_{-1}^{1} d\mu\,\mu\,\F_{ij} P_{\ell}
\end{equation}
Using the recurrence relation:
\begin{equation}
(\ell+1)P_{\ell+1}-(2\ell+1)\mu P_{\ell} + \ell P_{\ell-1} = 0
\end{equation}
we can express $\mu P_\ell(\mu)$ in terms of $P_{\ell+1}$ and $P_{\ell-1}$. We then get:
\begin{multline}
\frac{i^{\ell+1}k}{2} \int_{-1}^{1} d\mu\,\mu\,\F_{ij} P_{\ell} =\frac{i^{\ell+1}k}{2} \sum_{m=0}^\infty (-i)^{m} (2m+1)\F_{ij}^{(m)} \int_{-1}^{1} d\mu \left[\frac{(\ell+1)P_{\ell+1}+\ell P_{\ell-1}}{2\ell+1}\right] P_{m}=\\[0.7cm]
=\frac{i^{\ell+1}k}{2} \sum_{m=0}^\infty (-i)^{m} (2m+1)\F_{ij}^{(m)}\times \left[\frac{2(\ell+1)}{(2\ell+1)(2m+1)}\delta_{\ell+1,m} + \frac{2\ell}{(2\ell+1)(2m+1)}\delta_{\ell-1,m}\right]=\\[0.7cm]
=\frac{i^{\ell+1}k}{2\ell+1} \left[  (-i)^{\ell+1} (\ell+1) \F_{ij}^{(\ell+1)} + (-i)^{\ell-1} \ell \F_{ij}^{(\ell-1)} \right] =  \frac{k}{2\ell+1} \left[   (\ell+1) \F_{ij}^{(\ell+1)} -  \ell\, \F_{ij}^{(\ell-1)} \right].\\
\end{multline}
\paragraph{Third term.}
We need to calculate:
\begin{equation}
\frac{i^\ell}{2} \int_{-1}^{1} d\mu\,\left[2\dot h_{l m}\int_0^{2\pi} d\phi\, n^l n^m \left(n_i n_j-\frac{\delta_{ij}}{3}\right) \right]  P_{\ell}
\end{equation}
Let us start from the angular integration.  
It can be shown that, if $A_{ij}$ is a generic (three-dimensional) symmetric, transverse traceless tensor, then:
\begin{align}
&A_{lm}\int_0^{2\pi}d\phi\, n_l n_m=0\\
&A_{lm}\int_0^{2\pi}d\phi\, n_l n_m n_i n_j =\frac{\pi}{2}(\mu^2-1)^2 A_{ij}
\end{align}
So the term reduces to :
\begin{equation}
\frac{i^\ell}{2} \int_{-1}^{1} d\mu\,\left[2\dot h_{l m}\int_0^{2\pi} d\phi\, n^l n^m n_i n_j \right]  P_{\ell}(\mu) = \frac{\pi i^\ell}{2} \dot h_{ij }\int_{-1}^{1} d\mu\,(\mu^2-1)^2   P_{\ell}(\mu)= \frac{\pi i^\ell}{2} \dot h_{ij } \gamma^{(\ell)}
\end{equation}
where the $\gamma^{(\ell)}$ are related to the coefficients of the expansion in Legendre polynomials of the function $(\mu^2-1)^2 = (8/15) P_0 - (16/21) P_2 + (8/35) P_4$ and are given by:
\begin{equation}
\gamma^{(\ell)} = \left\{
\begin{array}{rc}
\displaystyle{\frac{16}{15}} & \ell = 0, \\[0.5cm]
\displaystyle{-\frac{32}{105}} & \ell = 2, \\[0.5cm]
\displaystyle{\frac{16}{315}} & \ell = 4, \\[0.5cm]
0 & \qquad\mathrm{otherwise.}
\end{array}
\right.
\end{equation}
Putting all together we finally get:
\begin{equation}
\dot \F_{ij}^{(\ell)}+\frac{k}{2\ell+1} \left[   (\ell+1) \F_{ij}^{(\ell+1)} -  \ell\, \F_{ij}^{(\ell-1)} \right] + \frac{\pi i^\ell}{2} \dot h_{ij } \gamma^{(\ell)} = 0,
\end{equation}



\paragraph{Note added in the arXiv version} After this paper was published, we became aware 
that the effect of sudden changes in the radiation energy density on the spectrum of GWs has been studied
in Ref. \cite{Watanabe:2006qe}. Moreover, in Ref. \cite{Mangilli:2008bw} the effects of free-streaming neutrinos have been computed up to second order in perturbation
theory.

\begin{acknowledgements}
This work has been developed in the framework of the CGW collaboration
(www.cgwcollaboration.it).
\end{acknowledgements}



\end{document}